\documentclass{emulateapj}
\usepackage{apjfonts}
\usepackage{epsfig}
\usepackage{graphicx}

\newcommand{\cubecm}{\ifmmode{~{\rm cm^{-3}}}\else{~cm$^{-3}$}\fi}
\newcommand{\kms}{\ifmmode{~{\rm km~s^{-1}}}\else{~km s$^{-1}$}\fi}
\newcommand{\lsim}{\lower0.3em\hbox{$\,\buildrel <\over\sim\,$}}
\newcommand{\gsim}{\lower0.3em\hbox{$\,\buildrel >\over\sim\,$}}

\newcommand{\rvir}{\ifmmode{{\rm r_{vir}}}\else{r$_{\rm{vir}}$}\fi}
\newcommand{\mvir}{\ifmmode{{\rm M_{vir}}}\else{M$_{\rm{vir}}$}\fi}
\newcommand{\beq}{\begin{equation}}
\newcommand{\eeq}{\end{equation}}

\shorttitle{Reionization and the $z=0$ Universe}
\shortauthors{ALVAREZ ET AL.}

\begin{document}

\title{Connecting Reionization to the Local Universe}

\author{Marcelo A. Alvarez, Michael Busha, Tom Abel, and Risa H. Wechsler}

\affil{Kavli Institute for Particle Astrophysics and
  Cosmology, Stanford University, Menlo Park, CA 94025}
\email{malvarez@slac.stanford.edu}

\begin{abstract}
  We present results of combined N-body and three-dimensional
  reionization calculations to determine the relationship between
  reionization history and local environment in a volume 1 Gpc
  $h^{-1}$ across and a resolution of about 1 Mpc.  We resolve the
  formation of about $2 \times 10^6$ halos of mass greater than $\sim
  10^{12} M_\odot$ at $z=0$, allowing us to determine the relationship
  between halo mass and reionization epoch for galaxies and clusters.
  For our fiducial reionization model, in which reionization begins at
  $z\sim 15$ and ends by $z\sim 6$, we find a strong bias for
  cluster-size halos to be in the regions which reionized first, at
  redshifts $10<z<15$. Consequently, material in clusters was
  reionized within relatively small regions, on the order of a few
  Mpc, implying that all clusters in our calculation were reionized by
  their own progenitors.  Milky Way mass halos were on average
  reionized later and by larger regions, with a distribution most similar
  to the global one, indicating that low mass halos are nearly uncorrelated
  with reionization when only their mass is taken as a prior.  On
  average, we find that most halos with mass less than $10^{13}
  M_\odot$ were reionized internally, while almost all halos with
  masses greater than $10^{14} M_\odot$ were reionized by their own
  progenitors.  We briefly discuss the implications of this work in
  light of the ``missing satellites'' problem and how this new
  approach may be extended further.
\end{abstract}


\section{INTRODUCTION}

The universe we observe at $z=0$ must bear the marks of reionization.
Reionization began when the first stars polluted the intergalactic
medium and created individual H~II regions \citep{alvarez/etal:2006,
abel/etal:2007,yoshida/etal:2007,wise/abel:2008}.  As
the first galaxies grew in abundance, the H~II regions became longer
lived, eventually containing perhaps tens of thousands of dwarf
galaxies, growing and merging until they overlapped, marking the end
of
reionization \citep{shapiro/giroux:1987,miralda-escude/etal:2000,gnedin:2000a,sokasian/etal:2001,nakamoto/etal:2001,ciardi/etal:2003,furlanetto/etal:2004,iliev/etal:2006,zahn/etal:2007,trac/cen:2007}. Observations
of high-redshift quasars imply that this process was 
complete by redshift $z\sim 6$ 
\citep{becker/etal:2001,fan/etal:2002,white/etal:2003,
willott/etal:2007}, while large-angle polarization measurements of
the cosmic microwave background constrain the duration of
reionization \citep{spergel/etal:2003,komatsu/etal:2008}.

During this time, the temperature of the 
intergalactic medium increased from a few to tens of thousands of
degrees, dramatically changing the evolution of gas as it responded to
the highly dynamic underlying dark matter potential. 

Low mass halos in ionized regions are less able cool, collapse, and form stars  
than those in neutral regions, due to the increase in the
cosmological Jeans mass when gas is ionized and photo-heated,
sometimes called ``Jeans mass filtering'' \citep[e.g.,][]{shapiro/etal:1994, 
thoul/weinberg:1996,gnedin:2000b,dijkstra/etal:2004,shapiro/etal:2004}.
This suppression of structure is one of the fundamental ways that
reionization can leave its imprint on subsequent structure
formation, even up until the present day.

Correlating reionization with the present-day environment may be the
key to the so-called ``missing satellite problem'' --- many more
satellite halos are predicted to form in CDM than are actually
observed as galaxies \citep{klypin/etal:1999}.  The leading
explanation --- an alternative to more exotic possibilities like
modifying dark matter or the amplitude of small-scale primordial
density fluctuations --- is that the UV background maintains the
intergalactic gas in a photo-heated state, preventing it from falling
into the shallow potential wells of the progenitors of the satellite
halos \citep[e.g.,][]{bullock/etal:2000,benson/etal:2002}. For
example, one might expect that regions that were reionized earlier
will have fewer luminous satellites than regions that were ionized
later.  However, biased regions, which are rich in early low-mass
galaxy formation, would have reionized {\em first}.  The latter effect
implies that early reionization would lead to more satellite galaxies,
while the former implies just the opposite.  Detailed
three-dimensional models are necessary in order to disentangle these
competing effects and quantify their dependence on the inevitable
assumptions that must be made when modeling reionization on such large
scales.

In this {\em Letter}, we present our first calculations to address the
correlation between reionization and local environment.  We take a
novel approach, combining N-body simulations 
with a ``semi-numerical'' algorithm
\citep{zahn/etal:2007,mesinger/furlanetto:2007}  
for calculating the reionization history for the simulation,
allowing us to achieve a higher dynamic range in
resolving the scales of reionization than has been possible until
now. We then report on the statistical correlations between halo
properties and their reionization epoch and environment.  We present
our hybrid N-body/semi-numerical method in \S 2, our results in \S 3,
and end with a discussion in \S 4.  Throughout, we assume a flat
universe with $\Omega_m=0.25$, $\sigma_8=0.8$, $n_s=1$,
$\Omega_b=0.04$, and $h=0.7$.  


\begin{figure*}
\begin{center}
\includegraphics[width=1.0\textwidth]{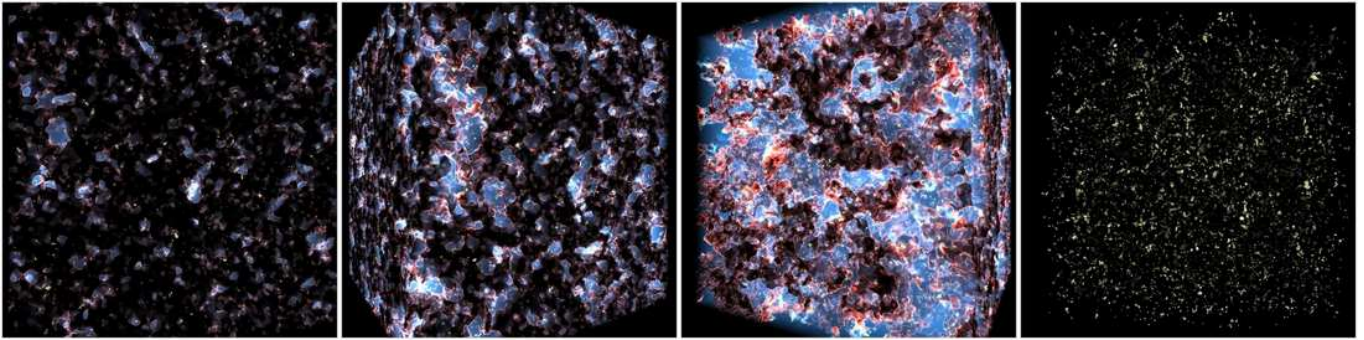}
\end{center}
\caption{Visualization of the progress of reionization in our 1
  Gpc$h^{-1}$ calculation. Redshifts $z=14$, $z=10$, $z=8$, and $z=6$ are
  shown from right to left. Ionized regions are
  blue and translucent, ionization fronts are red and white, and
  neutral regions are dark and opaque. A random sampling of 5 per cent
  (about 40,000) of all the halos at $z=0$ are shown in yellow.  Reionization is still quite inhomogeneous on 
these large scales, with large
  regions ionizing long before others.
\label{panels}
} 
\end{figure*}

\section{MODEL}

Our hybrid approach consists of two steps. First, we run an N-body
simulation of structure formation to determine the positions and
masses of halos at $z=0$.  We then calculate the reionization history
of the same volume in order to determine the reionization epoch of
each halo.\newpage

\subsection{N-body Dark Matter Simulations}

For our cosmological N-body simulations, we used the code GADGET-2
\citep{springel:2005}.  We simulated a periodic box 1 Gpc$/h$ on a side
with $1120^3$ N-body particles.  We did not include any gas dynamics,
a reasonable choice given that we are interested only in the
global properties of the dark matter halos, not the internal
properties of the baryonic component.  We used a comoving softening
length of 25 kpc, sufficient to resolve the formation of halos of
mass $\sim 10^{12}$~$M_\odot$.  At $z=0$ we use a friends-of-friends
halo finder with a linking length of 0.2 mean inter-particle spacings
to identify the halos. 

\subsection{Semi-numerical Reionization}

Our model is based on the analytical formalism first introduced by
\citet{furlanetto/etal:2004} and later extended to three dimensional realizations by
\citet{zahn/etal:2007}.  Its main assumption is that a region is fully
ionized if 
its collapse fraction, defined as the fraction of matter present in
halos above some minimum mass $M_{\rm min}$, is greater than some
threshold, 
\beq
\zeta f_{\rm coll} > 1.
\label{original}
\eeq  
This corresponds, for example, to the assumption that
$\zeta\dot{f}_{\rm coll}$ ionizing photons are released per atom per
unit time.  If recombinations are neglected, then equation
(\ref{original}) results by ensuring that the time-integrated number
of ionizing photons released is greater than the number of the atoms.  
Another interpretation of the efficiency factor $\zeta$ is that each
halo produces a spherical ionized region around it, the size of which
is directly proportional to its mass.  Thus, all the recombination,
and radiative transfer physics is absorbed into our choice of $\zeta$.
For example, $\zeta = (f_{esc}f_{*}N_{\gamma/b})/(1 + n_{rec})$
where $f_{esc}$ is the escape fraction of ionizing photons from each halo,
$f_{*}$ is the fraction of matter converted to stars within a halo, 
$N_{\gamma/b}$ is the number of ionizing photons produced in stars per
hydrogen atom, $n_{rec}$ is the average number of recombinations per
hydrogen atom during reionization \citep{furlanetto/etal:2004}. 

To apply this criterion for ``self-ionization'' to an actual
three-dimensional linear density field, we use the following relation
for the collapsed fraction within a spherical region of mass $m$ and
density contrast $\delta$ \citep{lacey/cole:1993}: 
\begin{equation}
f_{coll} = {\rm erfc}\left[\frac{\delta_{c}(z) - \delta_m}{\sqrt{2[\sigma^{2}(M_{\rm min}) - \sigma^{2}(m)]}}\right],
\label{extendedps}
\eeq
where $\sigma^{2}(m)$ is the mass variance over the scale $m$, 
$\delta_{c}(z)$ is the critical density for collapse, and $M_{\rm min}$ is
the minimum mass of halos to be counted in the collapsed fraction, i.e. the minimum mass of a halo capable of producing a significant amount of photoionizing radiation.
Note that the time dependence of the density field has been
taken into account in the critical density for collapse,
$\delta_c(z)=\delta_{c,0)}/D(z)$, where $D(z)$ is the linear growth 
factor, so that $\sigma(m)$ and $\delta_m$ are constant in time. 
As shown by \citet{furlanetto/etal:2004}, this results in a time and scale-dependent
``barrier" around each point, 
\begin{equation}
\delta_{m} \geq \delta_{x}(m, z) \equiv \delta_{c}(z) - \sqrt{2}\left[\sigma^{2}(M_{\rm min}) - \sigma^{2}(m)\right]^{1/2} \mbox{erf}^{-1}(1 - \zeta^{-1
}).
\label{deltamcondition}
\end{equation}
The mean density within a sphere around a given point, $\delta_m$,
must be greater than this barrier, $\delta_{x}(m, z)$, in order for
the point at the center to be ionized by that region.  A given point
is considered to be ionized when the condition in equation
\ref{deltamcondition} is met for {\em any} smoothing scale $m$, so
that
\beq
z_{reion}={\rm MIN}_m \left[D_0\left(\delta_m+\sqrt{2}\ {\rm erf}^{-1}(1-\zeta)
[\sigma^2(m_{\rm min})-\sigma^2(m)]\right)-1\right],
\eeq
where ${\rm MIN}_m$ indicates the minimum value over all smoothing
scales $m$. 

In practice, the outcome of this modeling is one value of $z_{reion}$
at each point on the grid, which characterizes the evolution of
reionization over all time.  The smoothing of the density field over
all scales can be accomplished through a fast Fourier transform
(FFT).  We also store the radius at which each point on the grid first
crossed the barrier, and associate it with it with the characteristic
size of the region containing the sources that first ionized the grid
point. 

Finally, to associate with each $z=0$ halo a reionization epoch and H~II
region size, we assign each of them a value that corresponds to the cell in
which its center of mass lies at present.  
Given that typical H~II
regions are tens of Mpc, the vast majority of halos in our
volume would not have had the required sustained peculiar velocities
in excess of $10^3$ km/s for $10$~Gyrs to have moved out of such 
a region. We thus expect our results to be robust for
most halos in the box, with the predictions being least accurate for
the few halos that are just on the verge of falling
into large galaxy clusters. Here, the reionization epochs 
could be overestimated, and H~II region sizes underestimated.
We set the parameters of the reionization model to have a
minimum halo mass of $M_{\rm min}=10^8 M_\odot$ and an efficiency
parameter $\zeta=10$.

\section{RESULTS}

Figure~\ref{panels} shows snapshots of the reionization
calculation at four different times for a 1024$^3$ grid.  Even on scales of 100
Mpc, reionization remains inhomogeneous, with the reionization
redshift of regions as large as tens of Mpc varying between $z_r\sim
15$ and $z_r \sim6$.  Although regions that were reionized first are at
the peaks of the underlying density field, there is not a one-to-one
correspondence between the mass of the $z=0$ halos and their
reionization epochs, since the halo and reionization barrier shapes
and amplitudes differ. 

\begin{figure}
\begin{center}
\includegraphics[width=0.3\textheight]{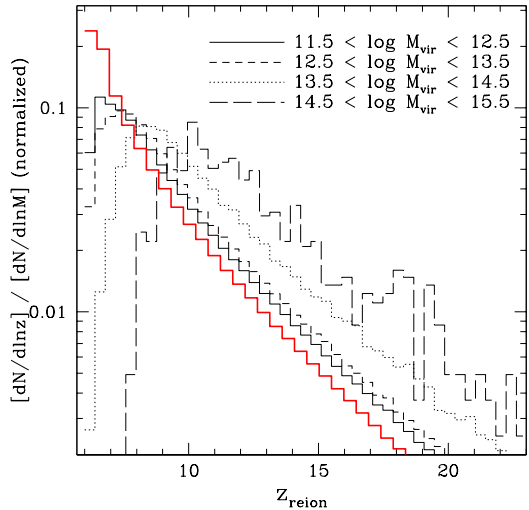}
\end{center}
\caption{Reionization redshift distributions, for the regions that
  became halos in different mass
  intervals (see labels) by redshift zero.  A uniform binning was used in redshift,
  with the same spacing for each of the mass ranges.  The low-mass
  distributions are very well sampled because of their large numbers,
  while the high-mass halos, especially at the cluster scale, become
  noisy, owing to their small number (only $~800$ even in a 1 Gpc$h^{-1}$
  box).  The thick red line shows the distribution for random points
  in the box. 
\label{binz_m}
}
\end{figure}

Figure~\ref{binz_m} showns the distribution of halo reionization
epochs, for several ranges of halo mass.  There is 
considerable spread in reionization redshifts in this model, ranging
over $6 < z_r < 15$.  The most massive halos are biased toward
higher values of $z_r$, peaking at $z\sim 10$, 8, and 7 for
masses $M\sim 10^{15}$, $10^{14}$, and $10^{13}~M_\odot$. The
distribution of the lowest mass halos, with masses $\sim
10^{12}~M_\odot$, does not have a well-defined peak, but rather
increases toward the lowest redshifts, peaking at the percolation
epoch at $z\sim 6$.  
This indicates that these lowest mass halos are
relatively unbiased with respect to the structure of reionization.

\begin{figure}[t]
\begin{center}
\includegraphics[width=0.3\textheight]{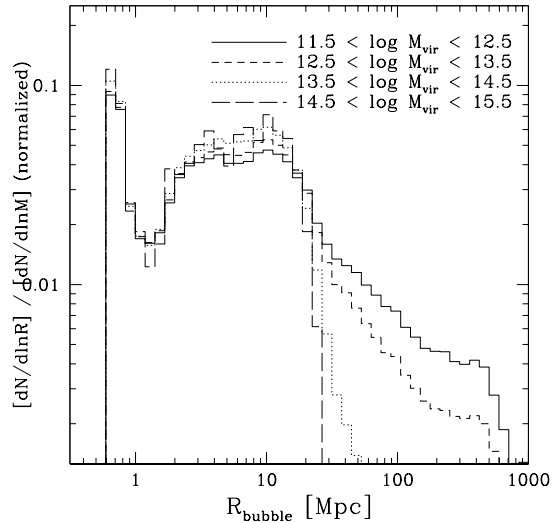}
\end{center}
\caption{H~II region sizes at reionization for the
  same mass bins as Figure~\ref{binz_m}.  The peak at low masses can be attributed to the grid resolution, given
  by a radius of $\sim 0.5$~Mpc. Increased
  resolution would reduce the effect, but on such
  small scales the role of recombinations and the
  stochasticity of the sources are likely to play an increasing role,
  effects we have not yet self-consistently included in these
  calculations.  The distribution at scales larger than $~\sim
  1$~Mpc are robust predictions of our model.
\label{binr_m}
}
\end{figure}

Figure \ref{binm_con} shows the 68 and 95 per
cent contours of the reionization redshift distribution for halos
binned by their mass.  The median value increases from $z_r\simeq 8$ for
$M_h=10^{12}~M_\odot$, to $z\simeq 12$ for $M_h=10^{15}~M_\odot$.  The
distributions have a long tail toward higher reionization values,
which is more pronounced for higher-mass halos.  Only 5 percent of
$10^{12}$-$M_\odot$ halos have $z_r>12$, while only 5 percent of
cluster scale halos have $z_r<8$.   Such a large spread in
reionization epochs at all masses implies that other halo properties,
such as merger history and local matter density, may be important in setting the reionization epoch for a specific halo
such as our own Milky Way.

The mass-dependent distributions of bubble sizes are shown in Figure \ref{binr_m}.
Lower mass halos largely form in regions with
larger H~II bubbles, since their sizes increase with time.
Interestingly, all of the roughly 800 cluster-mass halos in our sample
are associated with H~II region sizes less than 30 Mpc.  Only halos
below about $10^{13}~M_\odot$ have H~II regions sizes in excess of
100~Mpc, exceeding the mean free path for Lyman-limit systems and
approaching (and potentially exceeding) the size of the box, 1~Gpc$h^{-1}$.

\begin{figure}[b]
\begin{center}
\includegraphics[width=0.3\textheight]{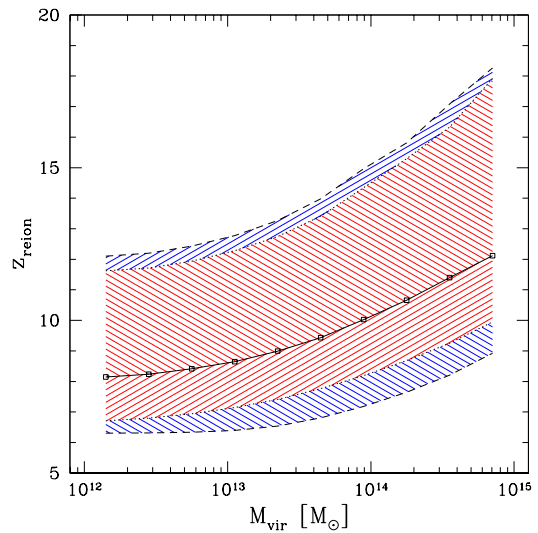}
\end{center}
\caption{Median (solid line) and spread in the values of the reionization redshift,
  $z_{\rm reion}$, as a function of halo mass.  Shaded contours indicate the
  68 per cent (red) and 95 per cent (blue) spread in the
  distribution. The higher the halo mass today the earlier its progenitors were likely
  reionized. 
\label{binm_con}
}
\end{figure}

\section{DISCUSSION}

Using large-volume and high-resolution coupled simulations of
reionization and halo formation, we have developed a new method for
connecting the $z=0$ distribution of halos to the reionization epoch.
We have found that, when only their mass is known, galaxy scale halos are
nearly uncorrelated with respect to reionization, with a distribution of H~II
region bubble sizes and reionization epochs that are roughly consistent with
having a random spatial distribution.  Higher mass halos, however, show
a much stronger correlation, with none of the cluster scale objects
having $z_r<8$ or $R_{\rm HII}>30$~Mpc.  

An important distinction is between {\em internal} and {\em external}
reionization.  Figure~\ref{schematic} describes these two
possibilities.  In the external reionization case, the 
halo material was ionized by sources in a region with
$R_{HII}\gg R_{\rm Lag}$, where $R_{\rm Lag}$ is the comoving volume
occupied by the mass of the halo at the cosmic mean density.  In
this case, most of the sources that ionized the material 
were not progenitors of the halo, and the
ionization front swept over the halo's progenitors quickly, leaving the
halo with a relatively uniform reionization epoch.  For internal
reionization, $R_{HII}<<R_{\rm Lag}$, and the halo's
reionization history is likely to be much more complex.  In general,
more massive halos were internally reionized, while less massive
ones were externally reionized.

Our definition is somewhat different from
previous definitions \citep[e.g.,][]{weinmann/etal:2007}, but we
believe our definition is best suited for the method used here.  In
our definition, halos are considered to be externally ionized if their
Lagrangian radius, defined by $M_{\rm halo}=4\pi\overline{\rho}R_{\rm
  Lag}^3/3$, is smaller than their H~II region radius.  For galaxy
scale objects, with Lagrangian radii of order 2 Mpc, it is clear from
Figure~\ref{binr_m} that most of these objects were externally
ionized.  Because our model does not resolve scales
below about a Mpc, however, it is difficult to determine how 
few galaxies were internally ionized.  More detailed
modeling of the small-scale structure on galactic scales, while still retaining the
large volume presented here, is therefore
necessary.   Our predictions for clusters are more
robust. For these objects, with Lagrangian radii of order
20~Mpc, all but a tiny handful are internally ionized.

Our results may have important implications for galaxy formation, and
in particular for the missing satellites problem.  Because we find
such a large spread in reionization epochs for Milky Way mass halos,
more information, such as larger-scale environment and accretion history, will be
necessary to determine the reionization epoch of our own halo, even
after the global reionization history is well constrained.
If the abundance of Galactic satellites is strongly dependent on the reionization epoch of
the galaxy, our results indicate that Milky Way mass halos would have
a large spread in the number of observable satellites.
We explore this issue in a companion paper (Busha et al., in preparation).

\begin{figure}[t!]
\begin{center}
\includegraphics[width=0.45\textwidth]{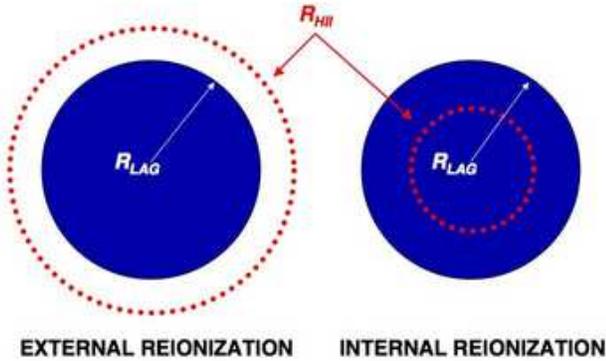}
\end{center}
\caption{Schematic diagram indicating the difference between external
  and internal reionization. $R_{HII}$ indicates the size of the
  reionizing bubble; $R_{Lag}$ indicates the Lagrangian radius
of the halo.  
\label{schematic}
}
\end{figure}

Our results may also have implications for the issue of galaxy
``assembly bias'', the idea that galaxy clustering may be dependent
on properties other than the mass of their host halos
\citep{wechsler/etal:2006, gao/white:2007, croton/etal:2007}. If the reionization epoch of
halos at a given mass is correlated with halo formation time, and if 
the reionization epoch affects any aspects of the galaxy population,
then assembly bias could be more important for such galaxies than it
is for their host halos.  Further study will be required to investigate such effects.

The approach we have presented here will serve as the foundation for
such more detailed future studies. These studies will investigate the
statistical correlations between present-day structure and
reionization, and will also incorporate detailed galaxy formation modeling.
Such improvements will allow for an investigation of the detailed
coupling between star formation histories and the local reionization history.

\acknowledgments
This work was partially supported by NASA ATFP grant NNX08AH26G and
NSF AST-0807312. RHW was supported by a Terman Fellowship at Stanford
University. We thank Louis Strigari for discussion about the missing
satellite problem.  MAA thanks Ilian Iliev and Piero Madau for helpful
discussions.  The Gpc simulation was performed on the Orange cluster
at SLAC as a part of the LasDamas project; MTB and RHW thank their
collaborators on the LasDamas project 
(http://lss.phy.vanderbilt.edu/lasdamas/)
for critical input. We are grateful for the continuous support from
the SLAC computational team.

  \bibliography{refs} \bibliographystyle{apj}

\begin{thebibliography}{35}
\expandafter\ifx\csname natexlab\endcsname\relax\def\natexlab#1{#1}\fi

\bibitem[{{Abel} {et~al.}(2007){Abel}, {Wise}, \& {Bryan}}]{abel/etal:2007}
{Abel}, T., {Wise}, J.~H., \& {Bryan}, G.~L. 2007, \apjl, 659, L87

\bibitem[{{Alvarez} {et~al.}(2006){Alvarez}, {Bromm}, \&
  {Shapiro}}]{alvarez/etal:2006}
{Alvarez}, M.~A., {Bromm}, V., \& {Shapiro}, P.~R. 2006, \apj, 639, 621

\bibitem[{{Becker}(2001)}]{becker/etal:2001}
{Becker}, R.~H. e.~a. 2001, \aj, 122, 2850

\bibitem[{{Benson} {et~al.}(2002){Benson}, {Frenk}, {Lacey}, {Baugh}, \&
  {Cole}}]{benson/etal:2002}
{Benson}, A.~J., {Frenk}, C.~S., {Lacey}, C.~G., {Baugh}, C.~M., \& {Cole}, S.
  2002, \mnras, 333, 177

\bibitem[{{Bullock} {et~al.}(2000){Bullock}, {Kravtsov}, \&
  {Weinberg}}]{bullock/etal:2000}
{Bullock}, J.~S., {Kravtsov}, A.~V., \& {Weinberg}, D.~H. 2000, \apj, 539, 517

\bibitem[{{Ciardi} {et~al.}(2003){Ciardi}, {Stoehr}, \&
  {White}}]{ciardi/etal:2003}
{Ciardi}, B., {Stoehr}, F., \& {White}, S.~D.~M. 2003, \mnras, 343, 1101

\bibitem[{{Croton} {et~al.}(2007){Croton}, {Gao}, \&
  {White}}]{croton/etal:2007}
{Croton}, D.~J., {Gao}, L., \& {White}, S.~D.~M. 2007, \mnras, 374, 1303

\bibitem[{{Dijkstra} {et~al.}(2004){Dijkstra}, {Haiman}, {Rees}, \&
  {Weinberg}}]{dijkstra/etal:2004}
{Dijkstra}, M., {Haiman}, Z., {Rees}, M.~J., \& {Weinberg}, D.~H. 2004, \apj,
  601, 666

\bibitem[{{Fan} {et~al.}(2002){Fan}, {Narayanan}, {Strauss}, {White}, {Becker},
  {Pentericci}, \& {Rix}}]{fan/etal:2002}
{Fan}, X., {Narayanan}, V.~K., {Strauss}, M.~A., {White}, R.~L., {Becker},
  R.~H., {Pentericci}, L., \& {Rix}, H.-W. 2002, \aj, 123, 1247

\bibitem[{{Furlanetto} {et~al.}(2004){Furlanetto}, {Zaldarriaga}, \&
  {Hernquist}}]{furlanetto/etal:2004}
{Furlanetto}, S.~R., {Zaldarriaga}, M., \& {Hernquist}, L. 2004, \apj, 613, 1

\bibitem[{{Gao} \& {White}(2007)}]{gao/white:2007}
{Gao}, L. \& {White}, S.~D.~M. 2007, \mnras, 377, L5

\bibitem[{{Gnedin}(2000{\natexlab{a}})}]{gnedin:2000a}
{Gnedin}, N.~Y. 2000{\natexlab{a}}, \apj, 535, 530

\bibitem[{{Gnedin}(2000{\natexlab{b}})}]{gnedin:2000b}
---. 2000{\natexlab{b}}, \apj, 542, 535

\bibitem[{{Iliev} {et~al.}(2006){Iliev}, {Mellema}, {Pen}, {Merz}, {Shapiro},
  \& {Alvarez}}]{iliev/etal:2006}
{Iliev}, I.~T., {Mellema}, G., {Pen}, U.-L., {Merz}, H., {Shapiro}, P.~R., \&
  {Alvarez}, M.~A. 2006, \mnras, 369, 1625

\bibitem[{{Klypin} {et~al.}(1999){Klypin}, {Kravtsov}, {Valenzuela}, \&
  {Prada}}]{klypin/etal:1999}
{Klypin}, A., {Kravtsov}, A.~V., {Valenzuela}, O., \& {Prada}, F. 1999, \apj,
  522, 82

\bibitem[{{Komatsu}(2008)}]{komatsu/etal:2008}
{Komatsu}, E. e.~a. 2008, ArXiv e-prints

\bibitem[{{Lacey} \& {Cole}(1993)}]{lacey/cole:1993}
{Lacey}, C. \& {Cole}, S. 1993, \mnras, 262, 627

\bibitem[{{Mesinger} \& {Furlanetto}(2007)}]{mesinger/furlanetto:2007}
{Mesinger}, A. \& {Furlanetto}, S. 2007, \apj, 669, 663

\bibitem[{{Miralda-Escud{\' e}} {et~al.}(2000){Miralda-Escud{\' e}},
  {Haehnelt}, \& {Rees}}]{miralda-escude/etal:2000}
{Miralda-Escud{\' e}}, J., {Haehnelt}, M., \& {Rees}, M.~J. 2000, \apj, 530, 1

\bibitem[{{Nakamoto} {et~al.}(2001){Nakamoto}, {Umemura}, \&
  {Susa}}]{nakamoto/etal:2001}
{Nakamoto}, T., {Umemura}, M., \& {Susa}, H. 2001, \mnras, 321, 593

\bibitem[{{Shapiro} \& {Giroux}(1987)}]{shapiro/giroux:1987}
{Shapiro}, P.~R. \& {Giroux}, M.~L. 1987, \apj, 321, L107

\bibitem[{{Shapiro} {et~al.}(1994){Shapiro}, {Giroux}, \&
  {Babul}}]{shapiro/etal:1994}
{Shapiro}, P.~R., {Giroux}, M.~L., \& {Babul}, A. 1994, \apj, 427, 25

\bibitem[{{Shapiro} {et~al.}(2004){Shapiro}, {Iliev}, \&
  {Raga}}]{shapiro/etal:2004}
{Shapiro}, P.~R., {Iliev}, I.~T., \& {Raga}, A.~C. 2004, \mnras, 348, 753

\bibitem[{{Sokasian} {et~al.}(2001){Sokasian}, {Abel}, \&
  {Hernquist}}]{sokasian/etal:2001}
{Sokasian}, A., {Abel}, T., \& {Hernquist}, L.~E. 2001, New Astronomy, 6, 359

\bibitem[{{Spergel}(2003)}]{spergel/etal:2003}
{Spergel}, D.~N. e.~a. 2003, \apjs, 148, 175

\bibitem[{{Springel}(2005)}]{springel:2005}
{Springel}, V. 2005, \mnras, 364, 1105

\bibitem[{{Thoul} \& {Weinberg}(1996)}]{thoul/weinberg:1996}
{Thoul}, A.~A. \& {Weinberg}, D.~H. 1996, \apj, 465, 608

\bibitem[{{Trac} \& {Cen}(2007)}]{trac/cen:2007}
{Trac}, H. \& {Cen}, R. 2007, \apj, 671, 1

\bibitem[{{Wechsler} {et~al.}(2006){Wechsler}, {Zentner}, {Bullock},
  {Kravtsov}, \& {Allgood}}]{wechsler/etal:2006}
{Wechsler}, R.~H., {Zentner}, A.~R., {Bullock}, J.~S., {Kravtsov}, A.~V., \&
  {Allgood}, B. 2006, \apj, 652, 71

\bibitem[{{Weinmann} {et~al.}(2007){Weinmann}, {Macci{\`o}}, {Iliev},
  {Mellema}, \& {Moore}}]{weinmann/etal:2007}
{Weinmann}, S.~M., {Macci{\`o}}, A.~V., {Iliev}, I.~T., {Mellema}, G., \&
  {Moore}, B. 2007, \mnras, 381, 367

\bibitem[{{White} {et~al.}(2003){White}, {Becker}, {Fan}, \&
  {Strauss}}]{white/etal:2003}
{White}, R.~L., {Becker}, R.~H., {Fan}, X., \& {Strauss}, M.~A. 2003, \aj, 126,
  1

\bibitem[{{Willott}(2007)}]{willott/etal:2007}
{Willott}, C.~J. e.~a. 2007, \aj, 134, 2435

\bibitem[{{Wise} \& {Abel}(2008)}]{wise/abel:2008}
{Wise}, J.~H. \& {Abel}, T. 2008, \apj, 684, 1

\bibitem[{{Yoshida} {et~al.}(2007){Yoshida}, {Oh}, {Kitayama}, \&
  {Hernquist}}]{yoshida/etal:2007}
{Yoshida}, N., {Oh}, S.~P., {Kitayama}, T., \& {Hernquist}, L. 2007, \apj, 663,
  687

\bibitem[{{Zahn} {et~al.}(2007){Zahn}, {Lidz}, {McQuinn}, {Dutta}, {Hernquist},
  {Zaldarriaga}, \& {Furlanetto}}]{zahn/etal:2007}
{Zahn}, O., {Lidz}, A., {McQuinn}, M., {Dutta}, S., {Hernquist}, L.,
  {Zaldarriaga}, M., \& {Furlanetto}, S.~R. 2007, \apj, 654, 12

\end{thebibliography}

\end{document}